\documentclass[conference]{IEEEtran}
\IEEEoverridecommandlockouts
\usepackage{comment}
\usepackage{verbatim}
\usepackage{booktabs}
\usepackage{cite}
\usepackage{amsmath}
\usepackage{multirow}
\usepackage{amsmath,amssymb,amsfonts}
\usepackage{algorithmic}
\usepackage{graphicx}
\usepackage{textcomp}
\usepackage[linesnumbered,ruled,vlined]{algorithm2e}
\usepackage{romannum}
\usepackage{xcolor}
\usepackage{float}

\def\BibTeX{{\rm B\kern-.05em{\sc i\kern-.025em b}\kern-.08em
    T\kern-.1667em\lower.7ex\hbox{E}\kern-.125emX}}
\begin{document}
\vspace{-4mm}
\title{Generative AI-driven Semantic Communication Framework for NextG Wireless Network \\

% \thanks{This work was supported by the Institute of Information and Communications Technology Planning and Evaluation (IITP) Grant funded by the Korea Government (MSIT) (Artificial Intelligence Innovation Hub) under Grant 2021-0-02068 and in part by the National Research Foundation of Korea (NRF) grant funded by the Korea government (MSIT) (No. 2020R1A4A1018607) *Dr. CS Hong is the corresponding author.}
}
\vspace{-6mm}
\author{\IEEEauthorblockN{Avi Deb Raha, Md. Shirajum Munir\textsuperscript{1, 2}, Apurba Adhikary\textsuperscript{1}, Yu Qiao\textsuperscript{3} and Choong Seon Hong\textsuperscript{1}*}
	\IEEEauthorblockA{\textsuperscript{1}\textit{Department of Computer Science and Engineering, Kyung Hee University, Yongin-si 17104, Republic of Korea}\\ \textsuperscript{2}\textit{School of Cybarsecurity, Old Dominion University, Norfolk, VA 23529, USA } \\ \textsuperscript{3}\textit{Department of Artificial Intelligence, Kyung Hee University, Yongin-si 17104, Republic of Korea}\\
	E-mail: avi@khu.ac.kr, mmunir@odu.edu, apurba@khu.ac.kr, qiaoyu@khu.ac.kr, cshong@khu.ac.kr}
}
\maketitle
%Next-generation networks must satisfy extensive and high-rate data requirements.
\vspace{-10mm}
\begin{abstract}
This work designs a novel semantic communication (SemCom) framework for the next-generation wireless network to tackle the challenges of unnecessary transmission of vast amounts that cause high bandwidth consumption, more latency, and experience with bad quality of services (QoS). In particular, these challenges hinder applications like intelligent transportation systems (ITS), metaverse, mixed reality, and the Internet of Everything, where real-time and efficient data transmission is paramount. Therefore, to reduce communication overhead and maintain the QoS of emerging applications such as metaverse, ITS, and digital twin creation, this work proposes a novel semantic communication framework. First, an intelligent semantic transmitter is designed to capture the meaningful information (e.g., the rode-side image in ITS) by designing a domain-specific Mobile Segment Anything Model (MSAM)-based mechanism to reduce the potential communication traffic while QoS remains intact. Second, the concept of generative AI is introduced for building the SemCom to reconstruct and denoise the received semantic data frame at the receiver end. In particular, the Generative Adversarial Network (GAN) mechanism is designed to maintain a superior quality reconstruction under different signal-to-noise (SNR) channel conditions. Finally, we have tested and evaluated the proposed semantic communication (SemCom) framework with the real-world 6G scenario of ITS; in particular, the base station equipped with an RGB camera and a mmWave phased array. Experimental results demonstrate the efficacy of the proposed SemCom framework by achieving high-quality reconstruction across various SNR channel conditions, resulting in $93.45\%$ data reduction in communication.

\end{abstract}

\begin{IEEEkeywords}
Terahertz (THz), millimeter-wave (mmWave), SAM, Semantic.
\end{IEEEkeywords}

\section{Introduction}
The seamlessly connected world offers new services such as autonomous driving, virtual reality (VR), extended reality (XR), and intelligent transportation systems (ITS). However, It poses novel obstacles to communication systems, such as resource scarcity, network traffic congestion, and scalable connectivity for edge intelligence. \cite{its}.
To perform the intelligent tasks for the ITS, such as creating digital twins \cite{vedio},  traffic congestion detection, monitoring suspicious vehicle positioning \cite{IoV}, and driving behavior characterization \cite{intrusive}, a massive number of images or videos need to be transmitted from the edges (i.e., video surveillance camera, the camera installed on road side unit) to the server (i.e., cloud, base station). This process consumes significant communication resources. Fortunately, semantic communication \cite{avi}, an emerging technology for next-generation communication systems like 6G and 5G-A, can intelligently transmit data by focusing on meaningful and task-oriented information extracted from images.

Compared to traditional methods, semantic communication offers considerable advantages, including lower latency, less bandwidth usage, reduced data transmission, and increased throughput \cite{bash}, which is necessary for transmitting images for the ITS. Several studies have been done in the domain of image transmission. The initial innovation in semantic communication was seen with deep joint source-channel coding (JSCC), specifically designed for image transmission \cite{djscc}. In \cite{robust}, the authors used a generative adversarial network (GAN) based approach for the robust transmission of images under noise. However, both approaches transfer a latent image representation rather than extracting semantic information. To focus on more semantic features, the authors of \cite{segmentedgan} used a semantic segmentation map to extract semantic features at the transmitter. The receiver can use these maps along with a style image to generate similar images using a Conditional GAN. While the images generated at the receiving end may have structural similarities, they do not ensure accurate identification of vehicles, persons, or other objects. This limitation applies broadly to image generation techniques, including both GANs and diffusion models, as the objects or persons depicted in the generated images are not identical to those originally transmitted. A semantic image transmission based on semantic segmentation is proposed in \cite{bash}. The authors have proposed to segment images semantically into regions of interest (ROI) and regions of non-interest (RONI). They have used two semantic encoders for the data compression for both regions. However, the process of segmentation poses a significant challenge, necessitating expert labeling for the categorization of the image into ROI and RONI. Additionally, using two models for ROI and RONI can cause complexity, posing constraints on the efficiency of this system. To address the issues, the authors of \cite{semsam} used the segment anything model to generate a mask of only the ROI.

However, the image encoder in SAM, with its hefty load of 632 million parameters, is ill-suited for applications like edge devices and ITS. The necessity for reduced latency and limited resource consumption in these tasks renders this high-volume model inappropriate. Moreover, the selection process of prompts for different tasks is essential as the SAM needs prompts to segment objects. Additionally, the RONI can contain important information such as the weather condition (i.e., rain, fog, cloud) and time (i.e., day, night). 

Given these complexities and the limitations of existing solutions, the central research question of this study is: 
\begin{itemize}
    \item \textit{How can we design an efficient semantic communication architecture for the transmission of sequential images in ITS that not only considers task-relevance and real-world constraints like low latency and limited resources, but also ensures the accurate identification of vehicles, persons or any other entity?}
\end{itemize}

To address the aforementioned question in this study, we proposed a novel architecture for the semantic transmission of sequential images for the ITS applications. The contributions are as follows:

\begin{figure}[htbp]
\centering
\includegraphics[width=7.5CM]{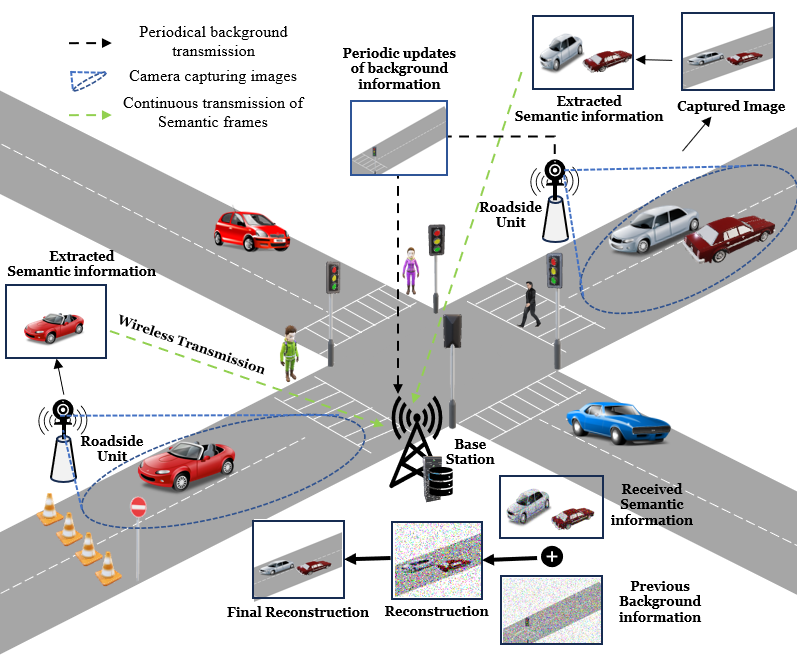}
\caption{System Model For Semantic Image Transmission in ITS.}
\label{System_Model}
\end{figure}
The following is a summary of the contributions:
\begin{itemize}
\item First, we propose a novel semantic communication framework which is designed explicitly for transmitting images or videos captured by video surveillance cameras on edge devices. In particular, our approach can transmit the prominent part of background information that capitalizes on the fact that the background of the surveillance camera images remains unchanged while leading to a significant reduction in data transmission.

\item Second, we design a semantic communication transmitter by  considering resource limitations inherent in edge devices and the need for low-latency transmission in real-world applications, we utilized a lightweight Mobile Segment Anything Model (MSAM) for essential semantic information extraction from the images. The designed semantic transmitter ensures efficient, effective information extraction without overburdening the devices.

%Instead of using proximity measures for the global model, we propose an algorithm in the global model for summarizing the cache decisions of the local models. That certainly helps to reduce the compilation time of the whole process.
\item Third, we employ a Generative Adversarial Network (GAN) based approach at the receiver end to reconstruct and denoise the received semantic frames with the background frame. This strategy resulted in a superior quality reconstruction under different Signal to Noise (SNR) channel conditions.

\item Finally, the simulation results indicate that the proposed method adeptly transmits sequential images captured at the transmitter, achieving a data reduction rate of \(93.45\%\), all while preserving the integrity of the original content.

\end{itemize}
%\subsection{Maintaining the Integrity of the Specifications}
\section{System Model and Problem Description}
\subsection{System Model}\label{AA}
The system model consists of two key components: a semantic transmitter and an intelligent receiver.
\subsubsection{Transmitter and The Process of Semantic Extraction}\label{AA}
Consider a transmitter $T_x$ equipped with an RGB camera that is recording video or capturing images. The transmitter in this system model can be a road side unit, traffic surveillance cameras, public surveillance cameras located in communal areas such as parks, streets, or business districts, or even stationary unmanned aerial vehicles (UAVs). In general, the transmitter can be anything that needs to transmit a set of sequential images $\mathcal{Q} = \left\{{Q[t_1],Q[t_2],\dots,Q[t_i],\dots,Q[t_T]}\right\}$ that has a almost static background. The information contained by these images can be divided into static information which is background of the image and dynamic or semantic information. For example, a RSU with a RGB camera takes images in a specific direction as shown in fig \ref{System_Model}. The background of the image is static as the structures of the background (i.e., road, building) remains almost same in the sequence of images. However, the images contain dynamic information such as vehicles, pedestrians, etc, that changes with the sequential frames. We consider these information as semantic information as these changable objects are the most important information contained by the image. The $Q[t_i]^{th}$ image therefore can be represented by the following equation:
\begin{figure}[htbp]
\centerline{\includegraphics[height=5cm, width=9cm]{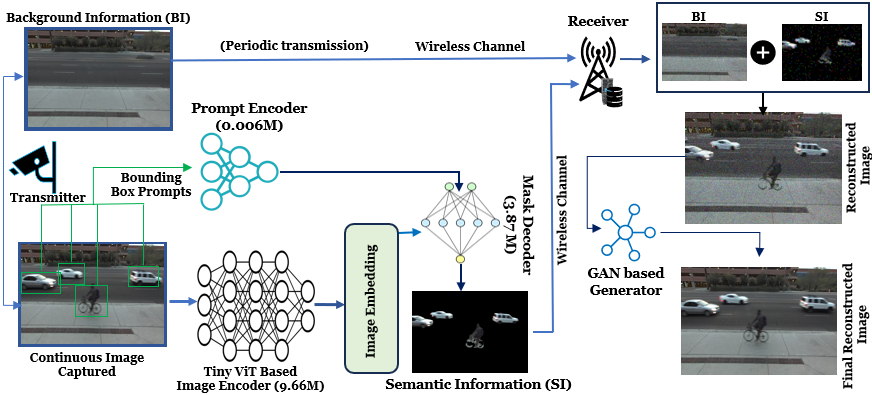}}
\caption{Step by step deep learning-based solution approach.}
\label{Solution approach}
\end{figure}
\begin{equation}
\textbf{Q}[t_i]=\textbf{B}[t_i]+\textbf{K}[t_i]\label{eq1}
\end{equation}
Here, $\textbf{Q}[t_i]$ is the $t_i^{th}$ image, $\textbf{B}[t_i]$ is the background information of the $t_i^{th}$ image and $\textbf{K}[t_i]$ is the semantic information of the image $\textbf{Q}[t_i]$. Now the next frame of the image can be represented as, 
\begin{equation}
\textbf{Q}[t_{i+1}]=\textbf{B}[t_i]+\textbf{K}[t_{i+1}]\label{eq2}
\end{equation}
Here, the $\textbf{B}[t_i]$ remains same as the background of the image does not change frequently. However, the $\textbf{B}[t_i]$ only needs very few periodical updates in comparison to the semantic information. For example, if the weather (i.e., rain, cloud) or light in the background (i.e., morning, evening, night) changes the  $\textbf{B}[t_i]$ needs an update. For simplicity, we consider a constant periodic update of the background information $\textbf{B}[t_i]$. The transmitted semantic signal of the transmitter can be represented as, 
\begin{equation}
\textbf{x}[t_i]=\mathcal{C}_t(\mathcal{S}(\mathbf{Q}[t_i]))\label{eq3}
\end{equation}
Here, $\mathcal{C}_t(.)$ represents the channel encoder and $\mathcal{S}(.)$ represents the semantic information extractor from the image $\mathbf{Q}[t_i]$.
The periodical background information signal can be represented as,
\vspace{-4mm}
\begin{equation}
\textbf{z}[\theta_j]=\mathcal{C}_t(\mathbf{B}[t_j])\label{eq4}
\end{equation}
The transmitter will transmit the signal $\textbf{x}$ for every frame that it captured or recorded. On the other hand, the transmitter will transmit the background information $\textbf{z}$ after a period $\theta$.
\subsubsection{Receiver}\label{AA} The receiver receives the semantic information and background information transmitted by the transmitter. The received signals can be represented as\cite{djscc},
\begin{equation}
\textbf{y}_i=\mathbf{H.x}[t_i]+\mathbf{n}\label{eq5}
\end{equation}
\begin{equation}
\textbf{w}_j=\mathbf{H.z}[\theta_j]+\mathbf{n}\label{eq6}
\end{equation}
Here, $\mathbf{H}$ represents the channel matrix and $\mathbf{n}$ represents the  additive white Gaussian noise. the decoded signals can be represented as,
\vspace{-4mm}
\begin{equation}
\hat{\mathbf{x}}[t_i]=\mathcal{C}_r(\mathbf{y}_i)\label{eq7}
\end{equation}
\vspace{-4mm}
\begin{equation}
\hat{\mathbf{z}}[\theta_j]=\mathcal{C}_r(\mathbf{w}_j)\label{eq8}
\end{equation}
Here, $\mathcal{C}_r(.)$ represents the channel decoder. The reconstructed image can be represented as. 
\begin{equation}
\widehat{\textbf{Q}}[t_i]=\mathcal{G}_r(\hat{\mathbf{z}}[\theta_j]\oplus\hat{\mathbf{x}}[t_i])\label{eq9}
\end{equation}
Here $\oplus$ is the image composition operation and $\mathcal{G}_r$ represents a semantic image reconstructor that do combined image denoising as well as image reconstruction task. The image reconstruction task is needed as the semantic extractor $\mathcal{S}(.)$ may fail to extract some portion of the semantic information from the image.

\subsection{Quality Measures}\label{AA}
We employ a triad of distortion measurement parameters to evaluate the efficacy of our image reconstruction, namely Peak Signal-to-Noise Ratio (PSNR)~\cite{gain}, Multi-Scale Structural Similarity (MS-SSIM)~\cite{robust}, and Visual Information Fidelity (VIF)~\cite{vif}.

\subsubsection{Peak Signal-to-Noise Ratio (PSNR)}
PSNR allows us to calculate the differences at the granular pixel level between the original and reconstructed images. This metric is crucial for evaluating the fidelity of the reconstructed image in representing the original image without noise or distortion. The PSNR can be defined as follows \cite{gain},
\begin{equation}
\text{PSNR}(\textbf{Q}[t_i], \widehat{\textbf{Q}}[t_i]) = 10 \cdot \log_{10} \left( \frac{{\text{MAX}_I^2}}{{\text{MSE}(\textbf{Q}[t_i], \widehat{\textbf{Q}}[t_i])}} \right)
\end{equation}
where \(\text{MAX}_I\) - Maximum possible pixel value of the image and \(\text{MSE}(\textbf{Q}[t_i], \widehat{\textbf{Q}}[t_i])\) represents the mean squared error between the original and reconstructed images.

\subsubsection{Multi-Scale Structural Similarity (MS-SSIM)}
MS-SSIM quantifies the congruence in the multi-scale structures of the two images, extending the capabilities of the original SSIM metric \cite{gain2} by incorporating multiple scales. This is particularly important for ensuring that important features and textures are preserved across different scales as different downstream task need different level of important features. The MS-SSIM which extends the capabilities of the original SSIM  can be defined as \cite{robust},
\begin{equation}
\text{MS-SSIM}(\textbf{Q}[t_i], \widehat{\textbf{Q}}[t_i]) = \prod_{i=1}^{N} [\text{SSIM}(\textbf{Q}[t_i]^{(i)}, \widehat{\textbf{Q}}[t_i]^{(i)})]^{\alpha_i}
\end{equation}
Here,  \(N\) and \(\alpha_i\)  Number of scales and  Weighting factors for each scale.

\subsubsection{Visual Information Fidelity (VIF)}
VIF serves to assess the statistical correlation between the test and reference images, offering a measure of visual fidelity based on natural scene statistics. This is crucial for understanding how well the reconstructed image captures the nuances of the original image's content. The VIF can be represented as \cite{vif}, 
\begin{equation}
\text{VIF}(\textbf{Q}[t_i], \widehat{\textbf{Q}}[t_i]) = \frac{{\sigma_{\textbf{Q}[t_i], \widehat{\textbf{Q}}[t_i]}^2}}{{\sigma_{\textbf{Q}[t_i]}^2 + \sigma_{\widehat{\textbf{Q}}[t_i]}^2}}
\end{equation}
where, \(\sigma\) and \(\sigma_{\textbf{Q}[t_i], \widehat{\textbf{Q}}[t_i]}\) represents Standard deviation and  Cross-covariance between the original and reconstructed images respectively.

\section{Proposed Framework}
The proposed framework is divided into two sections: transmitter and receiver. The detail figure of the proposed architecture is given in figure\ref{Solution approach}. The components and their working procedure is given as follows:

\subsubsection{Transmitter} The transmitter takes images as input. Let $\mathbf{x} \in {\mathbf{R}^{C,W,H}}$ is an image. Here, $C,W,H$ represents the channel, height and width of the image respectively. We need to extract the semantic information from the image $\mathbf{x}$. To extract the semantic information we have used the Mobile Segment Anything Model (MSAM) \cite{MSAM}. The MSAM is a light weight version of the Segment Anything Model (SAM) \cite{SAM} which is developed by using knowledge distillation technique. SAM uses a heavy image encoder (ViT-H) that has $632$M parameters, which is highly time and resource consuming. Therefore it is not applicable where the resource is limited and latency is highly required. The MSAM uses a decoupled distillation process to distill the knowledge from the SAM encoder. The MSAM's image encoder is only $5.78$M which makes it more suitable for the transmitter. Like the SAM, the MSAM takes image and a prompt as input and gives segmented masks of the objects that is denoted by prompts. The prompt of the MSAM can be anything from the set $P_\textrm{prompt} \in \{P_\textrm{BoundingBox},P_\textrm{point},P_\textrm{mask},P_\textrm{text}\}$. We use bounding box as the prompt to get the segmented masks or the semantic information from the image $\textbf{Q}[t_i]$. Therefore the semantic information of the image $\textbf{Q}[t_i]$ can be represented as, 
\begin{equation}
\mathcal{S}(\mathbf{Q}[t_i])=\zeta(\mathbf{Q}[t_i],\phi_{t_i}^\mathcal{S}), \label{eq10}
\end{equation}
Here, $\zeta$ is the MSAM model and $\phi_{t_i}^\mathcal{S}$ is the bounding box prompts on the semantic objects $\mathcal{S}$ in image frame $\mathbf{Q}[t_i]$.
For detecting the bounding boxes $\phi_{t_i}^\mathcal{S}$, we have used an open-set object detection method named Grounding Dino (GD) \cite{dino}. The use of open-set object detection is motivated by its ability to address a limitation in traditional object detection. In the traditional approach, models are trained to recognize a fixed set of object classes and their performance is evaluated based on these specific classes. For instance, a model may be trained to detect cars, dogs, and humans, and subsequently, its testing involves only these objects. In contrast, open-set object detection is designed to handle scenarios where the object classes encountered during the testing phase can differ from, or even exceed, those covered in the training phase.  Furthermore, GD is capable of detecting a wide range of objects based on human inputs such as texts, which could include category names or descriptive phrases of the semantic information. Therefore, using GD adds scalability to the proposed system model.
The bounding box prompts for image $\mathcal{Q}[t_i]$ using the GD can be represented as the following,
\begin{equation}
\phi_{t_i}^\mathcal{S}=\Upsilon (\xi_\mathcal{S}^t,\mathbf{Q}[t_i]) \label{eq11}
\end{equation}
Here $\Upsilon$ is the GD model which takes input text prompt $\xi_\mathcal{S}^t$ and an image $\mathbf{Q}[t_i]$ and returns the bounding box prompts of semantic objects. $\xi_\mathcal{S}^t$ is a text prompt represents the semantic information $\mathcal{S}(.)$ that is needed from the image $\mathbf{Q}[t_i]$. For example, if we need all the persons and cars from the image $\mathbf{Q}[t_i]$, the prompt $\xi_\mathcal{S}^t$ will be $\xi_\mathcal{S}^t$= $" person. car"$. The semantic information can be rewrite as, 
\begin{equation}
\mathcal{S}(\mathbf{Q}[t_i])=\zeta(\mathbf{Q}[t_i], \Upsilon (\xi_\mathcal{S}^t,\mathbf{Q}[t_i])), \label{eq12}
\end{equation}
Algorithm 1 represents the semantic information and background information transmitting procedure.
\begin{algorithm}[!t]
	\caption{Proposed algorithm for the Transmitter}
	\label{alg1}
	\begin{algorithmic}[1]
		\renewcommand{\algorithmicrequire}{\textbf{Input: }}
		\renewcommand{\algorithmicensure}{\textbf{Output:}}
		\REQUIRE Set of image frames $\mathcal{Q}$.
		\ENSURE  $\mathbf{x}[t]$, $\mathbf{z}[\theta]$ and input text prompt $\xi_\mathcal{S}^t$.
        \\ \textbf{Initialization}: $\zeta$, $\Upsilon$, $\theta, \delta$, $\psi$ $\mathbf{B}$.
        % \STATE Divide $\mathcal{D}$ into $\mathcal{D}_{\textrm{train}}$ and $\mathcal{D}_{\textrm{test}}$
		\FOR{each $\mathbf{Q}[t_i] \in \mathcal{Q}$}
        \STATE \textbf{Find Bonding Boxes:} Compute $\phi_{t_i}^\mathcal{S}$ using Equation (\ref{eq11}).
        \IF{$\phi_{t_i}^\mathcal{S}$ == NULL}
        \STATE \textbf{Background:} update $\mathbf{B}[\theta]=\mathbf{Q}[t_i]$
        % \STATE \textbf{Increment:} Increment $\theta$ by one ($\theta=\theta+1$)
        \ENDIF
        \STATE \textbf{Extract Semantics:} Compute $\mathcal{S}(\mathbf{Q}[t_i])$ using Equation (\ref{eq10}).
		\STATE \textbf{Calculate:} Compute $\mathbf{x}[t_i]$ using Equation (\ref{eq3}).
            \STATE \textbf{Transmit:} Transmit $\mathbf{x}[t_i]$.
            \STATE \textbf{Increment:} Increment $\delta$ by one ($\delta=\delta+1$).
            \IF{$\delta==\psi$}
            \STATE \textbf{Calculate:} Compute $\mathbf{z}[\theta]$ using Equation (\ref{eq4}).
            \STATE \textbf{Transmit:} Transmit $\mathbf{z}[\theta]$.
            \STATE Reset $\delta$ to zero ($\delta=0$)
            \ENDIF
        \ENDFOR
	\end{algorithmic} 
\end{algorithm}
\subsubsection{Receiver} The receiver receives the signals $y$ and $w$ and recovers the frames $\hat{\mathbf{x}}[t_i]$ and $\hat{\mathbf{z}}[\theta]$ respectively. The received $\hat{\mathbf{x}}[t_i]$ is a masked image which only contains the semantic information. Therefore, before combining the $\hat{\mathbf{x}}[t_i]$ and $\hat{\mathbf{z}}[\theta]$, the corresponding parts of $\hat{\mathbf{z}}[\theta]$ should be set to zero where $\hat{\mathbf{x}}[t_i]$ is non-zero. This is to ensure that the semantic information in $\hat{\mathbf{x}}[t_i]$ does not conflict with the original background information in $\hat{\mathbf{z}}[\theta]$. Then the composed image can be represented as,
\begin{equation}
\mathbf{Q}[t_i^\varsigma] = \hat{\mathbf{z}}[\theta_j]\oplus\hat{\mathbf{x}}[t_i]\label{eq13}
\end{equation}
Next, the composed image needs to be reconstructed as it has wireless channel noise and some missing pixel that is caused by the light weight MSAM model. We have used the pix2pix GAN\cite{pix2pix} for a joint reconstruction and denoising the channel noise. In the training process of the GAN the generator takes  $\mathbf{Q}[t_i^\varsigma]$ as input image and try to map this input image to the ground truth image  $\mathbf{Q}[t_i]$. The adversarial loss of the generator $\mathcal{G}_r$ can be represented as\cite{pix2pix},
\vspace{-2mm}
\begin{equation}
L_{\text{GAN}}(\mathcal{G}_r,\mathcal{D}_r) = \mathbb{E}_{\mathbf{Q}[t_i^\varsigma],\mathbf{Q}[t_i]}\left[\log(\mathcal{D}_r(\mathbf{Q}[t_i],\mathcal{G}_r(\mathbf{Q}[t_i^\varsigma])))\right] \label{eq14}
\end{equation}
Here, $\mathcal{G}_r(\mathbf{Q}[t_i^\varsigma])$ represents the output images or reconstructed images, $\mathcal{G}_r$ and $\mathcal{D}_r$ represents the generator and discriminator respectively and $\mathcal{D}_r(\mathbf{Q}[t_i],\mathcal{G}_r(\mathbf{Q}[t_i^\varsigma]))$ is the prediction of the discriminator of the reconstructed image $\mathbf{Q}[t_i^\varsigma]$ with the original image $\mathbf{Q}[t_i]$. Next, to encourage the output image of the generator $\mathcal{G}_r(\mathbf{Q}[t_i^\varsigma])$ to resemble the target image $\mathcal{Q}[t]$ on a pixel-by-pixel level the following L1 loss is added to the adversarial loss $L_{\text{GAN}}(\mathcal{G}_r,\mathcal{D}_r)$. The L1 loss can be written as, 
\begin{equation}
L_{L1}(\mathcal{G}_r) = \mathbb{E}_{\mathbf{Q}[t_i^\varsigma],\mathbf{Q}[t_i]}[||\mathbf{Q}[t_i] - G(\mathbf{Q}[t_i^\varsigma])||_1]
\label{eq15}
\end{equation}
Therefore, the total loss function of the generator $\mathcal{G}_r$ can be written as\cite{pix2pix},
\vspace{-2mm}
\begin{equation}
L(\mathcal{G}_r,\mathcal{D}_r) = L_{GAN}(\mathcal{G}_r,\mathcal{D}_r) + \Phi L_{L1}(\mathcal{G}_r)
\label{eq16}
\vspace{-2mm}
\end{equation}
The L1 loss $L_{L1}(\mathcal{G}_r)$ helps to maintain the fidelity of the denoised and reconstructed image $\mathbf{Q}[t_i^\varsigma]$ to the original image $\mathbf{Q}[t_i]$ 
and the adversarial loss $L_{\text{GAN}}(\mathcal{G}_r,\mathcal{D}_r)$ helps to enhance the perceptual quality of the generated images, making them more realistic and natural-looking. Therefore we have used the combination of both of them. For the discriminator $(\mathcal{D}_r)$, the loss can be defined as\cite{pix2pix}, 
\vspace{-2mm}
\begin{align}
L_{\text{GAN}}(\mathcal{D}_r,\mathcal{G}_r) = & \mathbb{E}_{\mathbf{Q}[t_i^\varsigma],\mathbf{Q}[t_i]}[\log(\mathcal{D}_r(\mathbf{Q}[t_i^\varsigma],\mathbf{Q}[t_i]))] \nonumber \\
& + \mathbb{E}_{\mathbf{Q}[t_i^\varsigma],\mathbf{Q}[t_i]}[\log(1 - \mathcal{D}_r(\mathbf{Q}[t_i],\mathcal{G}_r(\mathbf{Q}[t_i^\varsigma])))] 
\label{eq17}
\vspace{-2mm}
\end{align}
The goal if the discriminator $L_{\text{GAN}}(\mathcal{D}_r,\mathcal{G}_r)$ loss  to correctly classify real $\mathbf{Q}[t_i]$  and generated images $\mathcal{G}_r(\mathbf{Q}[t_i^\varsigma])$ by the generator. That is, it wants to assign high probability to real samples $\mathbf{Q}[t_i]$ and low probability to generated samples $\mathcal{G}_r(\mathbf{Q}[t_i^\varsigma])$ . By doing this the discriminator helps the generator to generate less noisy and more perfect images.

\begin{algorithm}[!t]
\caption{Proposed Algorithm for the Receiver}
\label{alg:math_enhanced_receiver}
\begin{algorithmic}[1]
\renewcommand{\algorithmicrequire}{\textbf{Input:}}
\renewcommand{\algorithmicensure}{\textbf{Output:}}
\REQUIRE Set of received image frames \( \hat{\mathbf{z}}[\theta], \hat{\mathbf{x}}[t] \) and GAN model \( \mathcal{G}_r \).
\ENSURE Denoised and complete image $\widehat{\textbf{Q}}[t_i]$.
\\ \textbf{Initialization}: Pretrained Generator \( \mathcal{G}_r \), \( \mathbf{\Phi} \), \( \mathbf{\Psi} \), \( \mathbf{\Omega} \), $\mathbf{\Gamma}$.
\FOR{each \( \hat{\mathbf{x}}[t] \)}
\STATE \textbf{Set Background}: \( \mathbf{\Phi} \leftarrow \hat{\mathbf{z}}[\theta] \)
\STATE \textbf{Set Semantic Frame}: \( \mathbf{\Psi} \leftarrow \hat{\mathbf{x}}[t] \)
\STATE \textbf{Creating a binary mask}: \( \mathbf{\Omega}_{i, j} \leftarrow \bigvee_{k=1}^{c} (\mathbf{\Psi}_{i, j, k} \neq 0) \) // perform a logical OR across all the channels for each pixel located at (i,j) in the background
\STATE \( \mathbf{\Phi}_{i, j, k} \leftarrow \mathbf{\Phi}_{i, j, k} \cdot (1 - \mathbf{\Omega}_{i, j}) \)  $ //  \text{If } \mathbf{\Omega}_{i, j} \text{ is True, then } \mathbf{\Phi}_{i, j, k} \text{ is set to 0 for all } k$
\STATE \( \mathbf{Q}[t_i^\varsigma] \leftarrow \mathbf{\Phi}_{i, j, k} + \hat{\mathbf{x}}[t]_{i, j, k}  \)
\IF{\( \hat{\mathbf{x}}[t] = \varnothing \)}
\STATE \textbf{Use Background:} \( \mathbf{\Gamma}\leftarrow \hat{\mathbf{z}}[\theta] \)
\ELSE
\STATE \textbf{Use Combined Image:} \( \mathbf{\Gamma} \leftarrow \mathbf{Q}[t_i^\varsigma] \)
\ENDIF
\STATE \textbf{Denoising and Reconstruction:} \( \text{Output}: \widehat{\textbf{Q}}[t_i] \leftarrow \mathcal{G}_r(\mathbf{\Gamma}) \)
\ENDFOR
\end{algorithmic}
\end{algorithm}

\section{Simulation Results and Analysis}
\vspace{-2mm}
We have considered scenario 13 and scenario 3 from the DeepSense 6G \cite{dataset} dataset. 
These scenarios mainly include a fixed base station equipped with an RGB camera and a mmWave phased array. The camera is employed to obtain RGB pictures with a resolution of 960×540, operating at a fundamental frame rate of 30 frames per second (fps). However, the MSAM method with the GD model method can be applied to any dataset as the GD model is a open set object detection technique and the  MSAM has a zero shot learning capability. With the combination of this two we can extract semantic information from any dataset that matches the description of the system model.

% For the simulation process, we have considered the traffic concepts which are taken from TSRD dataset\cite{TSRD}. The dataset contains images of different traffic signs such as, \textit{left turn ahead}, \textit{red light signal}, \textit{no parking}, etc. The simulation parameters are shown in table \ref{tab:table1}. 
\begin{figure}[htbp]
\centerline{\includegraphics[width=5.5cm]{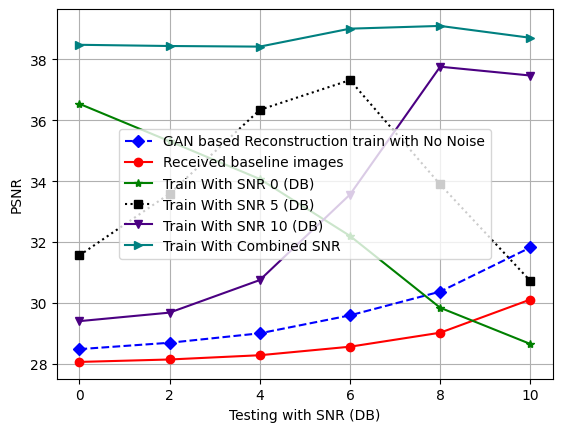}}
\vspace{-4mm}
\caption{Comparison of the PSNR score among different training setup.}
\label{PSNR}
\end{figure}

\begin{figure}[htbp]
\centerline{\includegraphics[width=5.5cm]{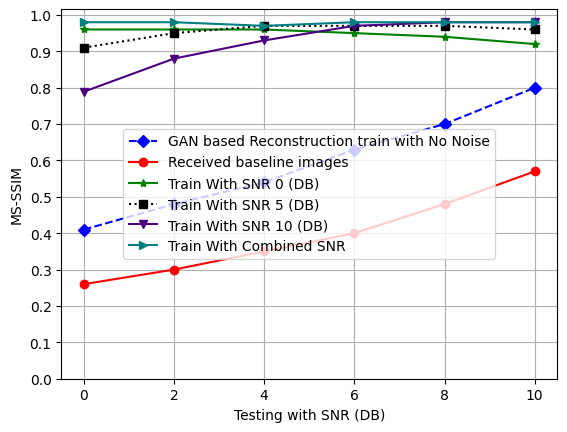}}
\caption{Comparison of the MS-SSIM score among different training setup.}
\vspace{-4mm}
\label{MSSIM}
\end{figure}

\begin{figure}[htbp]
\centerline{\includegraphics[width=5.5cm]{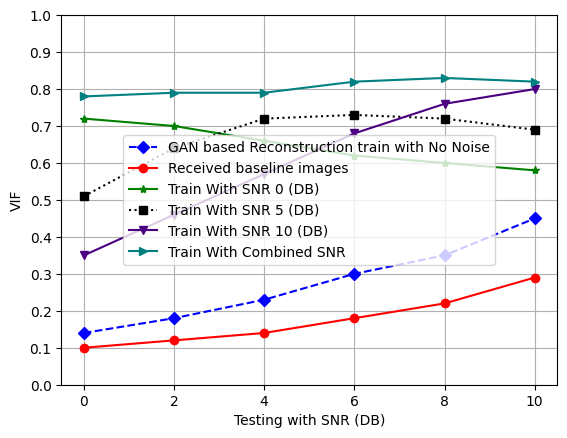}}
\vspace{-4mm}
\caption{Comparison of the VIF score among different training setup.}
\label{VIF}
\end{figure}

\begin{figure*}[htbp]
\centerline{\includegraphics[width=14cm]{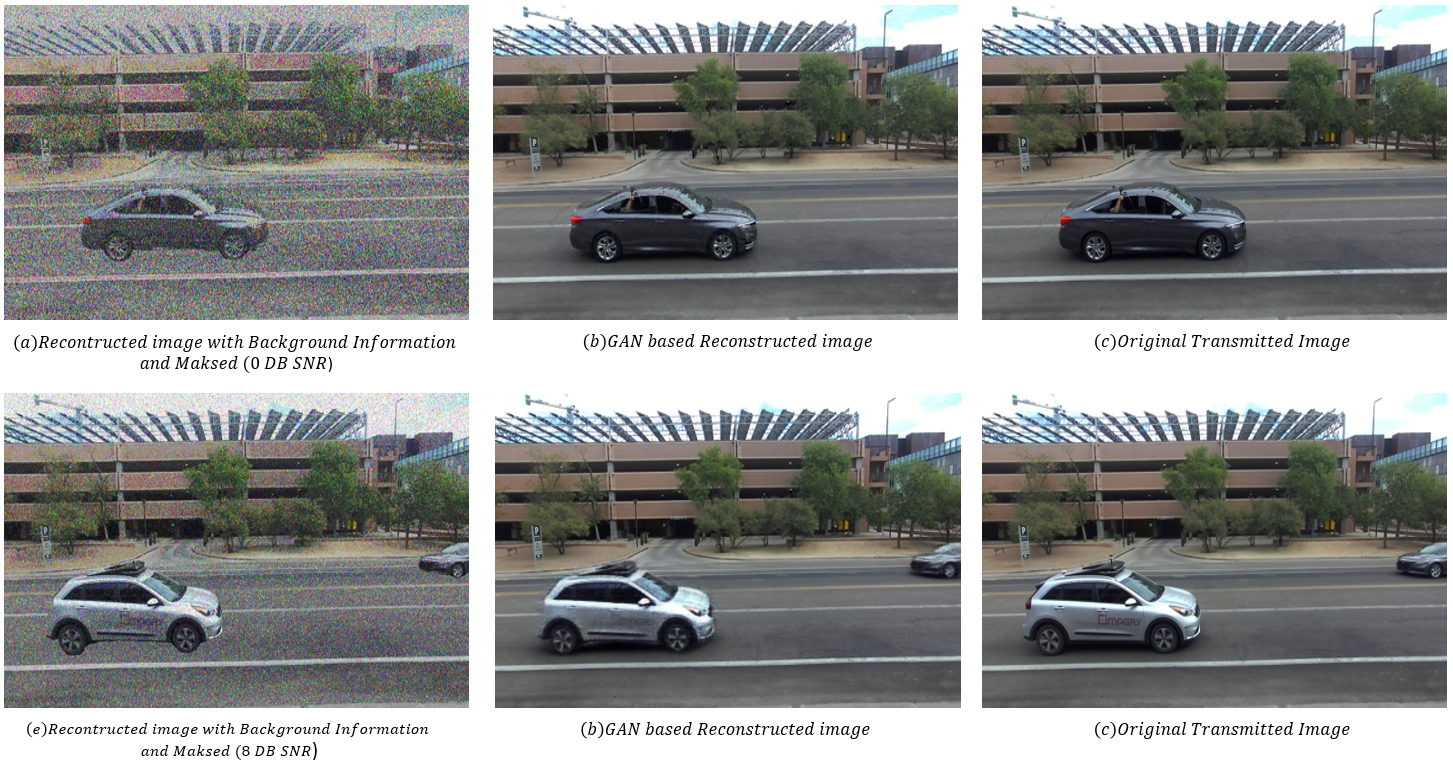}}
\vspace{-4mm}
\caption{Comparison of the Transmitted, Received and GAN based reconstructed images}
\label{comparedImages}
\end{figure*}

 \begin{figure}[htbp]
\centerline{\includegraphics[width=6cm]{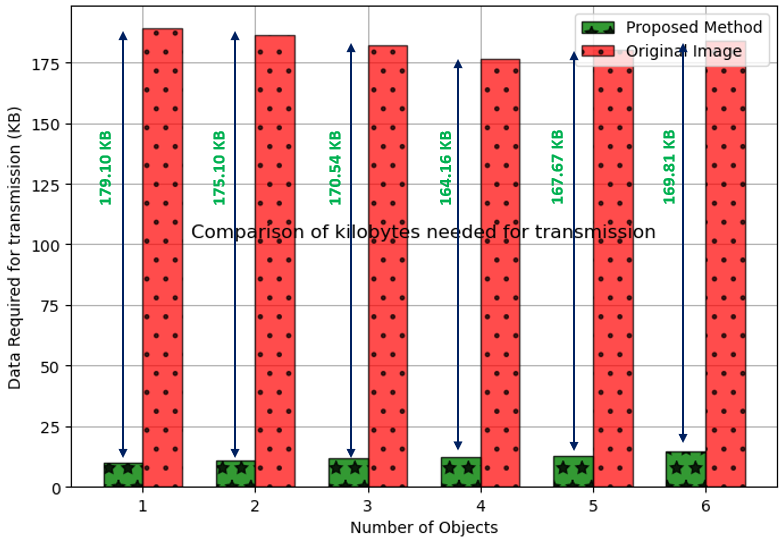}}
\vspace{-4mm}
\caption{Comparison of the data savings in JPEG format vs the proposed method.}
\label{savedData}
\end{figure}
\vspace{-4mm}
\subsection{Numerical Results}\label{AA}
We tested the proposed method in several signal to noise (SNR) conditions. More specifically, first we have trained the GAN based reconstruction in $\textrm{SNR}=0$ (DB), $\textrm{SNR}=5$ (DB) and $\textrm{SNR}=10$ (DB). The bit error probability (BEP) for the received images $\widehat{\textbf{Q}}[t_i]$ was $0.2854$, $0.1580$ and $0.0507$ for the $\textrm{SNR}=0$ (DB), $\textrm{SNR}=5$ (DB) and $\textrm{SNR}=10$ (DB) respectively. 
The figure referenced as \ref{PSNR} demonstrates the PSNR comparison among different training conditions. Interestingly, even though the GAN-based reconstruction trained without any noise consistently surpasses the baseline images in performance, it falls short when compared with the model trained with a diverse set of SNR levels (i.e., $0$ DB, $5$ DB, $10$ DB). This observation holds true irrespective of the SNR of the testing images. This pattern unravels a clear bias in the performance of the GAN-based reconstruction. Despite the high PSNR achieved by the GAN-based image construction, it is clear that the model performs best when the SNR of the testing and training images is closely matched. For instance, images trained with an SNR of $0$ DB achieved the highest PSNR of $36.54$ DB when tested with a $0$ DB SNR. A similar pattern is observed for both $5$ DB and $10$ DB training scenarios. Additionally, a clear degradation in performance is observed when the SNR of the testing image deviates significantly from the SNR of the training image. This suggests that the GAN based model is sensitive to the SNR level used during training, limiting its robustness and generalizability to a variety of conditions. Further insights into this behavior can be drawn from figures \ref{MSSIM} and \ref{VIF}. Figure \ref{MSSIM} represents the MS-SSIM comparison, and figure \ref{VIF} represents the VIF among the different SNR training levels in the GAN-based reconstruction.
To alleviate the observed bias, we trained the GAN using a combined set of SNR levels ($0$ DB, $5$ DB, $10$ DB) for image denoising and reconstruction. Figures \ref{PSNR}, \ref{MSSIM}, and \ref{VIF} collectively show that this training approach successfully mitigates the training bias, as it achieves the highest results and outperforms the other training SNR levels. This combined approach thus presents a more balanced and adaptable model for diverse SNR levels. Figure \ref{comparedImages} shows the reconstructed images. It can be seen from the reconstructed images that the GAN based reconstruction can successfully reconstruct and denoise the received images. Figure \ref{savedData} shows the comparison between the original images and the proposed method in terms of data transmission with respect to different number of objects in the image. The data indicates that the proposed method significantly reduces the amount of data transmitted compared to the original images. Specifically, from the figure it can be calculated that the average data reduction achieved by the proposed method is \(93.45\%\).

% \begin{table}[!t]
% \caption{Simulation Parameters\label{tab:table1}}
% \centering
% \begin{tabular}{|p{0.9cm}|c|p{0.7cm}||p{0.9cm}|c|p{0.7cm}|}
% \hline
% Method & Parameters & Values & Method & Parameters & Values \\
% \hline
% \multirow{6}{4em}{\hfil   PPO} & Batch Size & 64 & \multirow{6}{4em}{\hfil  CAE} & input shape & 28*28 \\ 
% & Discount Factor & 0.9 & & Epochs & 50 \\ 
% & Learning Rate & 0.003 & & Batch Size & 32\\ 
% & Clip Range & 0.02 & & Optimizer & adam\\ 
% & Episodes & 4e+6 & & learning rate & 0.001 \\ 
% & Epochs & 10 & & loss & mse\\
% \hline
% \end{tabular}
% \end{table}
% \renewcommand{\arraystretch}{1.5}
% \begin{table}[htbp]
%   \centering
%   \caption{Simulation Parameters\label{tab:table1}}
%     \begin{tabular}{cccc}
%     \toprule
%     \textbf{Parameters} & \textbf{LeNet5} & \textbf{VGG16} & \textbf{EfficientNetV2} \\
%     \midrule
%     \textbf{Batch} Size & 32 & 64 & 32 \\
%     \textbf{Epochs} & 30 & 50 & 30 \\
%     \textbf{learning rate}  & $10^-3$ & $10^-3$ & $10^-3$ \\
%     \textbf{LR Reduction Factor}  & 0.1 & 0.1 & 0.1 \\
%     \textbf{optimizer} & Adam & Adam & Adam \\
%     \textbf{Loss Function} & CrossEntropy & CrossEntropy & CrossEntropy \\
%     \bottomrule
%     \end{tabular}%
%   \label{tab:parameters}%
% \end{table}
\vspace{-2mm}
\section{Conclusion} 
% Our study demonstrates the transformative potential of Semantic Communication (SemCom) in future applications such as Intelligent Transport Systems (ITS), metaverse, mixed reality, and the Internet of Everything (IoE). We've innovatively reduced data transmission loads by focusing on transmitting meaningful information from edge device-captured videos and images. Leveraging the Mobile Segment Anything Model (MSAM), we extracted crucial semantic information, tackling the challenge of edge device resource limitations and low-latency requirements. The application of a Generative Adversarial Network (GAN) enabled us to reconstruct and denoise received frames, ensuring high-quality reconstructions across various Signal to Noise Ratio (SNR) channel conditions. Hence, our approach not only guarantees quality but also offers significant data reduction, paving the way for more efficient communication systems for next-generation applications.
In this work, we introduce a novel semantic communication framework that can efficiently transmit sequential images or videos while maintaining the original content unchanged. We have designed a semantic transmitter that can capture meaningful information using a domain-specific lightweight MSAM method. To reduce the transmission cost, the transmitter sends the background information periodically. The receiver receives the semantic information and merges it with the background information using a pix-to-pix GAN. The pix-to-pix GAN jointly reconstructs and denoises the images. The simulation result shows that the proposed framework can reduce up to $93.45\%$ of the communication cost while maintaining the original content. 
\bibliographystyle{IEEEtran}
\bibliography{references}

\end{document}